# A Primer on Bandwidth Parts in 5G New Radio

Xingqin Lin, Dongsheng Yu, Henning Wiemann

Ericsson

Emails: {xingqin.lin, dongsheng.yu, henning.wiemann}@ericsson.com

*Abstract*— **The fifth generation (5G) wireless access technology, known as New Radio (NR), features flexibility to support a variety of usage scenarios. One of the basic concepts in 5G NR is bandwidth part (BWP), which is, at a high level, a set of contiguous resource blocks configured inside a channel bandwidth. BWP spans across many 5G NR specifications developed by the 3rd Generation Partnership Project. Understanding how BWP operates is vital to understanding 5G NR. This article provides an overview of the essentials of BWP in the NR technical specifications. We describe fundamental BWP concepts, BWP configuration methods, and BWP switch mechanisms. We also discuss user equipment capabilities in terms of BWP support and share our thoughts on use cases of BWP for NR deployments.**

## I. INTRODUCTION

The 3rd Generation Partnership Project (3GPP) has developed a new radio-access technology known as New Radio (NR) in its Release 15, and continues to evolve NR to further improve performance and address new use cases in the fifth-generation (5G) era [1]. Compared to the previous generations of radio-access technologies, NR introduces many new features to support a wide range of services, devices, and deployments. In this article, we focus on one of the basic NR features – bandwidth part (BWP).

To develop a preliminary understanding of BWP, we first review the hierarchy of spectrum management in NR, which is illustrated in Figure 1. At a high level, NR defines frequency ranges (FRs). Currently, there are two FRs defined: The first is FR1 ranging from 410 – 7125 MHz [2], and the second is FR2 ranging from 24.25 – 52.6 GHz [3]. 3GPP further defines operating bands in each FR. An operating band is a frequency band associated with a certain set of radio frequency (RF) requirements. Bandwidths of different operating bands can vary from several MHz to a few GHz. Different operators may have different amounts of spectrum within an operating band. To accommodate diverse spectrum scenarios while limiting implementation complexity, NR supports a range of channel bandwidths from 5 – 400 MHz, where a channel bandwidth refers to the bandwidth of an NR carrier. The number of resource blocks (RBs) that may be configured in a channel bandwidth, known as transmission bandwidth configuration, shall meet the specified minimum guardband requirements [2][3]. Base station (BS) and user equipment (UE) can support different channel bandwidths. Like in Long-Term Evolution (LTE), a UE camps on and connects to a cell. The UE is made aware of the channel bandwidth of the cell. In addition to the cell bandwidth the network informs the UE about the position and width of a BWP. Loosely speaking, a BWP is hence a set of contiguous RBs configured inside a channel bandwidth. The width of a BWP may smaller than or equal to the cell bandwidth.

One motivation of introducing BWP in NR is to support UE bandwidth adaptation to help reduce device power consumption [4]. The main idea is that a UE may use a wide bandwidth when a large amount of data is scheduled, while being active on a narrow bandwidth for the remaining time. Another motivation is to support devices of different bandwidth capabilities by configuring the devices with different BWPs. A BS may support a very wide channel bandwidth which may not be supported by some UEs. BWP provides a mechanism to flexibly assign radio resources such that the signals for a UE are confined in a portion of BS channel bandwidth that the UE can support.

BWP, as a basic concept in NR, spans across different 3GPP specifications. Understanding how BWP operates is vital to developing a good knowledge of NR. A high-level introduction to BWP can be found in [4]. The white paper [5] provides further introduction to BWP concepts with a focus on UE power consumption. In contrast, the objective of this article is to delve into the detailed NR technical specifications to provide a complete overview of BWP design, while keeping the overall contents at a level accessible to an audience working in the wireless communications and networking communities. Besides, we take a network-centric approach and provide insights into NR deployments using BWPs.

The remainder of this article is organized as follows. In Section II, we introduce the basic concepts of BWPs. Then we describe how a network may configure BWPs in Section III and BWP switching mechanisms in Section IV. We discuss UE capabilities of supporting BWPs in Section V. Several use cases of BWPs for NR deployments are described in Section VI, followed by our concluding remarks in Section VII.

## II. BASIC CONCEPTS OF BANDWIDTH PARTS

### A. Fundamentals of Bandwidth Parts

NR defines scalable orthogonal frequency division multiplexing (OFDM) numerologies using subcarrier spacing (SCS) of $2^{\mu} \cdot 15$ kHz ($\mu = 0, 1, …, 4$) [6]. An RB consists of 12 consecutive subcarriers in the frequency domain. NR uses "Point A" as a common reference point for RB grids. The "Point A" is illustrated in Figure 1.

As illustrated in Figure 1, a BWP starts at a certain common RB and consists of a set of contiguous RBs with a given numerology (SCS and cyclic prefix) on a given carrier. For each serving cell of a UE, the network configures at least one downlink (DL) BWP (i.e., the initial DL BWP). The network may configure the UE with up to four DL BWPs, but only one

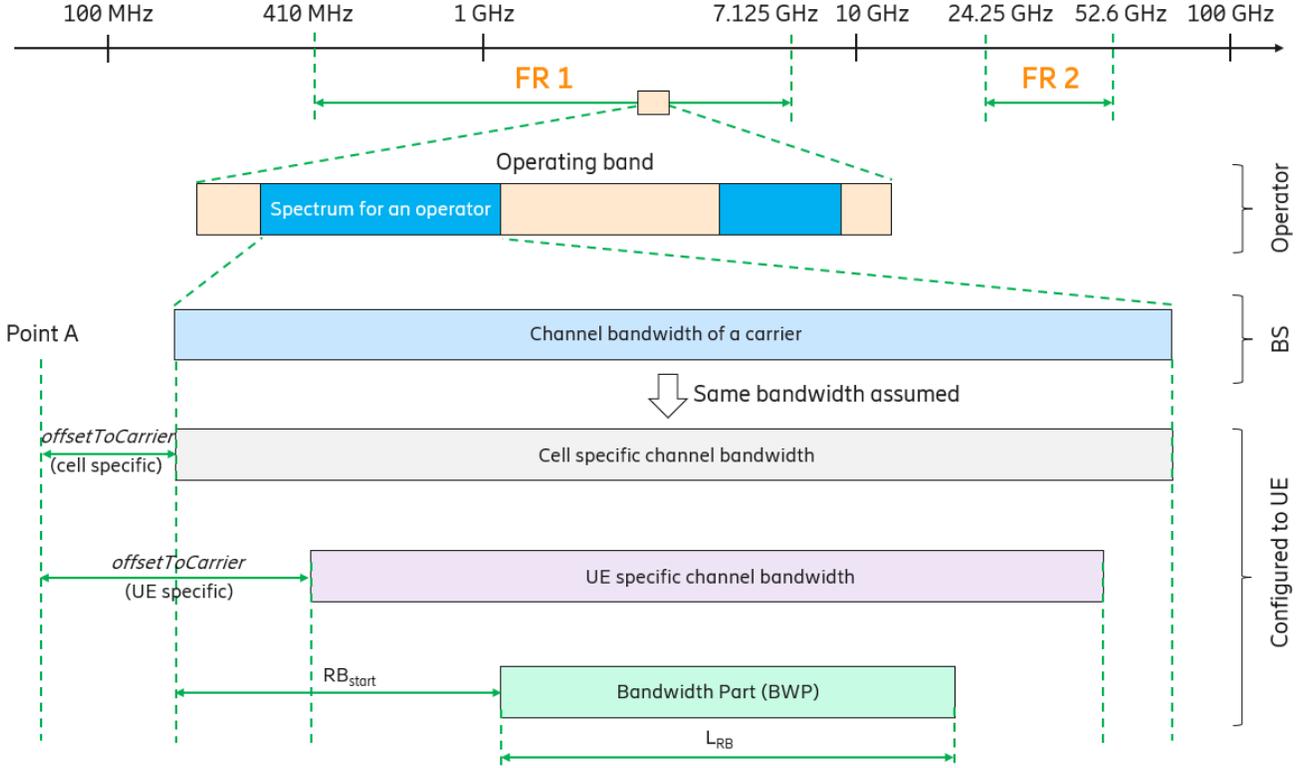

**Figure 1: An illustration of the 5G NR spectrum management and configuration**

DL BWP can be active at a given time. If the serving cell is configured with an uplink (UL), the network configures at least one UL BWP. Similar to the DL, the network may configure the UE with up to four UL BWPs, but only one UL BWP can be active at a given time. NR also supports a so-called supplementary UL (SUL), on which UL BWP(s) can be similarly configured as on a normal UL.

For paired spectrum, i.e., frequency division duplex (FDD), DL BWPs and UL BWPs are configured separately. For unpaired spectrum, i.e., time division duplex (TDD), a DL BWP is linked to an UL BWP when the indices of the two BWPs are the same. In this case, the paired DL BWP and UL BWP must share the same center frequency, but they can have different bandwidths.

In general, a UE only receives physical downlink shared channel (PDSCH), physical downlink control channel (PDCCH), or channel state information reference signal (CSI-RS) inside an active DL BWP. But the UE may need to perform radio resource management (RRM) measurements outside the active DL BWP via measurement gaps. Similarly, the UE only transmits physical uplink shared channel (PUSCH) or physical uplink control channel (PUCCH) inside an active UL BWP and, for an active serving cell, the UE does not transmit sounding reference signal (SRS) outside an active UL BWP.

### B. Bandwidth Parts Types

Activating an inactive BWP and deactivating an active BWP are called BWP switching to enforce that it is not possible to deactivate all BWPs or to activate more than one. For paired spectrum, DL BWPs and UL BWPs can be switched separately.

For unpaired spectrum, the paired DL BWP and UL BWP are switched together. The detailed BWP switching mechanisms are described in Section IV. In this subsection, we describe the types of BWPs that may be active at a given time.

**Initial DL/UL BWP:** The initial DL and UL BWPs are used at least for initial access before radio resource control (RRC) connection is established. An initial BWP has index zero and is referred to as BWP #0. During the initial access, the UE performs cell search based on synchronization signal block (SSB) composed of primary synchronization signal (PSS), secondary synchronization signal (SSS), and physical broadcast channel (PBCH). To access the system, the UE needs to further read system information block 1 (SIB1) which carries important information including the initial DL/UL BWP configuration. The SIB1 is transmitted on the PDSCH, which is scheduled by downlink control information (DCI) on the PDCCH using the control resource set with index zero (CORESET #0) [7] [8].

Before the UE reads the SIB1, the UE's initial DL BWP has the same frequency range and numerology as those of CORESET#0. After reading the SIB1, the UE follows the initial DL/UL BWP configuration in the SIB1 and uses them to carry out random-access procedure to request the setup of RRC connection. The network should configure the frequency domain location and bandwidth of the initial DL BWP in the SIB1 so that the initial DL BWP contains the entire CORESET #0 in the frequency domain.

**First active DL/UL BWP:** The first active DL and UL BWPs may be configured for a Special Cell (SpCell) or a secondary cell (SCell). In a master cell group (MCG), the SpCell refers to the primary cell (PCell) in which the UE performs the connection (re-)establishment procedure. In a secondary cell group (SCG), the SpCell refers to the primary SCG cell (PSCell) in which the UE performs random access for RRC (re-)configuration. An SCell provides additional radio resources on top of an SpCell in a cell group. The first active DL and UL BWPs are the active DL and UL BWPs upon RRC (re-)configuration for an SpCell or activation of an SCell.

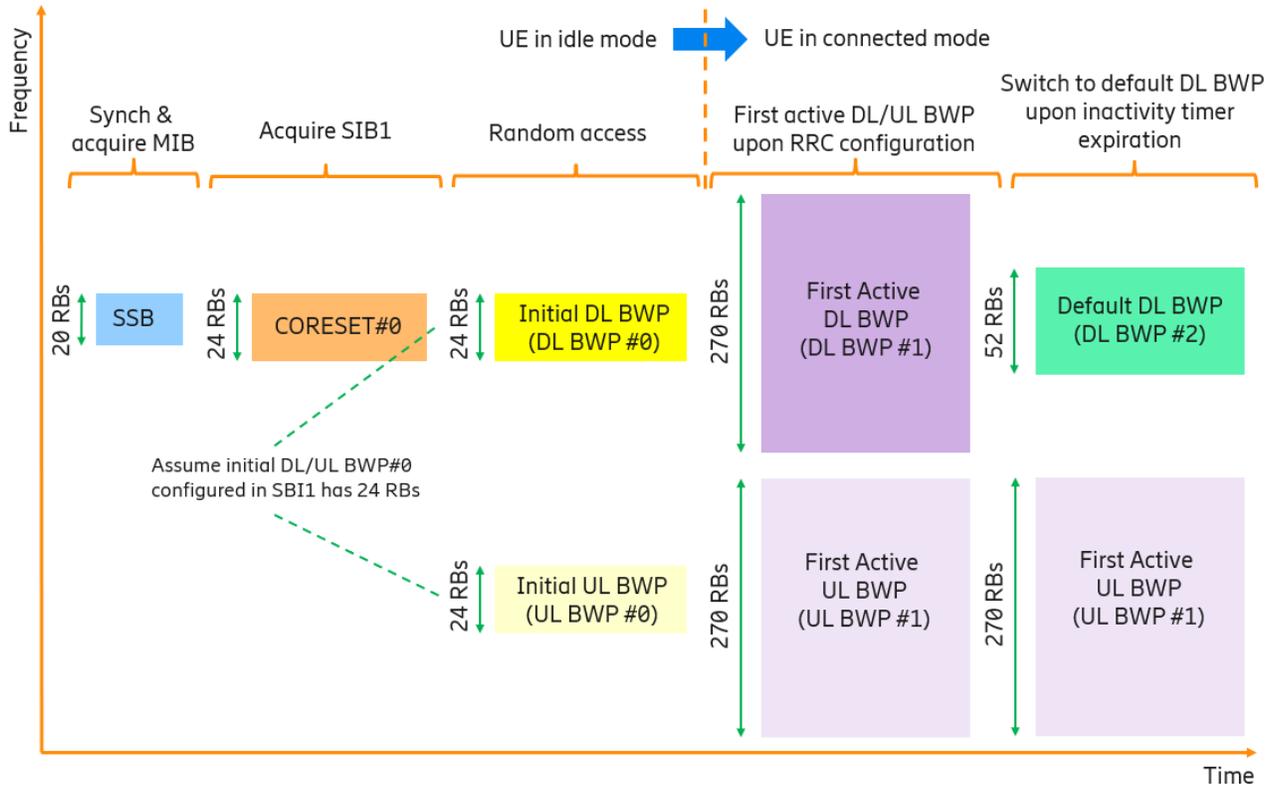

**Figure 2: An illustration of UE BWP adaptation from idle mode to connected mode.**

**Default BWP:** For a serving cell, the network may configure the UE with a BWP inactivity timer. The expiration of this timer may, for example, indicate that the UE has no scheduled transmission and reception for a while on the currently active BWP. Thus, the UE can switch its active BWP to a default BWP to save power. The default DL BWP can be configured. If not configured, the UE uses the initial DL BWP as the default DL BWP. For unpaired spectrum, when the UE switches its active DL BWP to the default DL BWP, the active UL BWP is switched accordingly since the BWP switching for TDD is common for both DL and UL.

Figure 2 provides an illustration of the aforementioned BWP types from a UE processing perspective. The UE first performs downlink synchronization and acquires PBCH based on 20-RB SSB. Assuming the CORESET#0 configured in the MIB has 24 RBs, the UE may assume that the initial DL BWP is 24 RBs wide and proceeds to acquire SIB1, which in this example also configures 24 RBs for both initial DL and UL BWPs. The UE then performs random-access procedure with the small initial DL and UL BWPs. After the random access, the UE reports that it is capable of supporting multiple BWPs. With dedicated RRC signaling, the network configures the UE with large DL/UL BWP #1 (270 RBs), small DL/UL BWP#2 (52 RBs), and BWP inactivity timer. The network sets the large DL/UL BWP #1 as the first active DL/UL BWP, and the small DL BWP #2 as the default DL BWP. Upon RRC configuration, the first active DL and UL BWPs (i.e., DL/UL BWP #1) become activated and are used for scheduling a large amount of data. After that, the UE does not have traffic demand and has no scheduled transmission. As a result, the BWP inactivity timer expires, upon which the UE switches its active DL BWP to the default DL BWP (i.e., DL BWP #2). Note that the active UL BWP does not need to switch to UL BWP #2, because Figure 2 illustrates an FDD system in which DL and UL BWPs are switched separately.

## III. BANDWIDTH PARTS CONFIGURATIONS

### A. Configuration of a Bandwidth Part with a Non-Zero Index

In this subsection, we discuss how to configure a BWP with a non-zero index. A DL/UL BWP with a non-zero index is a non-initial DL/UL BWP (recall that the index zero is reserved for initial DL/UL BWP), and is configured in addition to the initial DL/UL BWP.

The DL/UL BWP configurations are divided into common and dedicated parameters. The BWP-common parameters are cell specific, implying that the network needs to ensure that the corresponding parameters are appropriately aligned across the UEs. The BWP-dedicated parameters are UE specific.

The BWP-common parameters for a DL BWP with a non-zero index include basic cell-specific BWP parameters (frequency domain location, bandwidth, SCS, and cyclic prefix of this BWP) and additional cell-specific parameters for the PDCCH and PDSCH of this DL BWP. The BWP-dedicated parameters for a DL BWP with a non-zero index include UE-specific parameters for the PDCCH, PDSCH, semi-persistent scheduling, and radio link monitoring configurations of this DL BWP. The BWP-common parameters for an UL BWP with a non-zero index include basic BWP parameters and cell-specific parameters for the random access, PUCCH, and PUSCH of this UL BWP. The BWP-dedicated parameters for an UL BWP with a non-zero index include UE-specific parameters for the PUCCH, PUSCH, SRS, configured grant, and beam failure recovery configurations of this UL BWP.

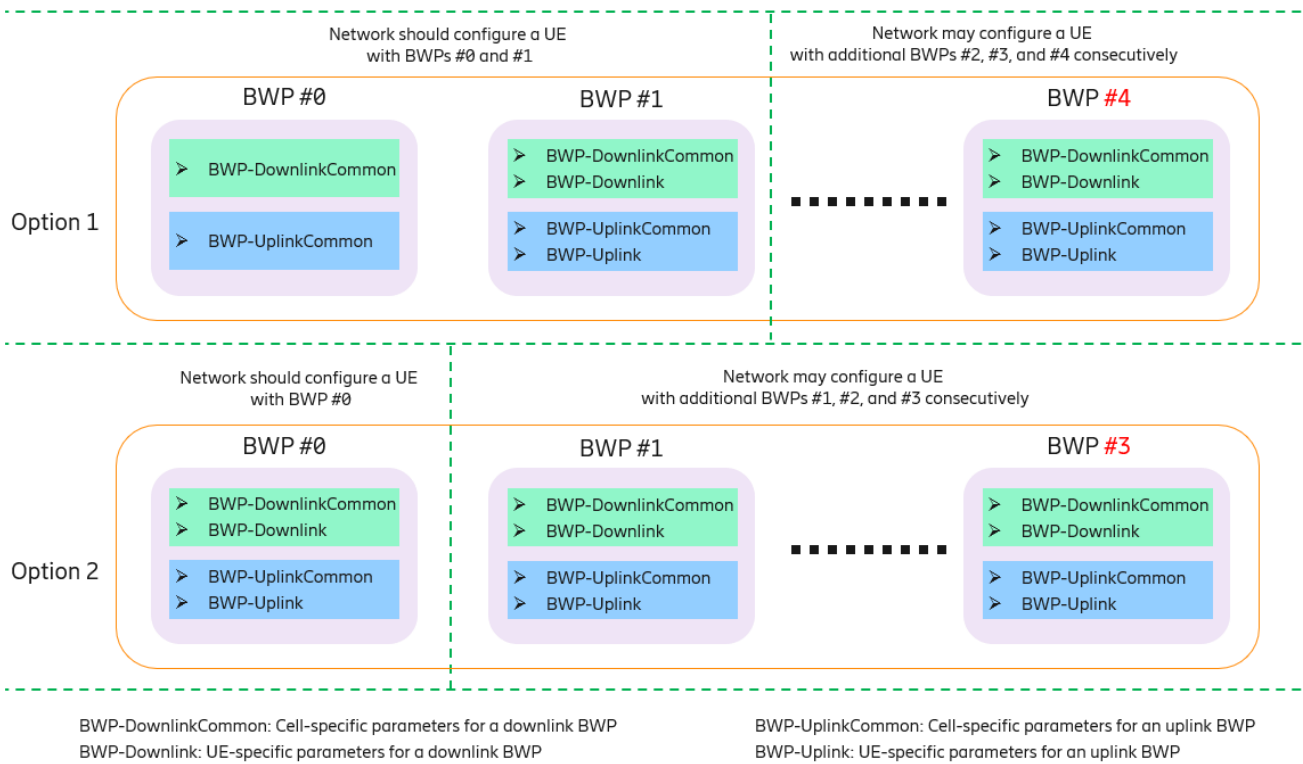

**Figure 3: An illustration of the 5G NR bandwidth parts configuration options**

### B. Configuration of a Bandwidth Part with Index Zero

There are two options for configuring a BWP with index zero (i.e. the initial BWP):

- Option 1: Configure the BWP #0 with cell-specific parameters only;
- Option 2: Configure the BWP #0 with both cell-specific and UE-specific parameters.

The DL/UL BWP #0 configured by Option 1 does not have the dedicated parameters and thus has limited functionality. In this case, the DL/UL BWP #0 mainly plays a temporary role and is used by the UE, for example, during the initial-access procedure. To set up a fully operational connection, the network should also configure the UE with an additional full-featured DL/UL BWP equipped with both cell-specific and UE-specific parameters.

The DL/UL BWP #0 configured by Option 2 is a full-featured BWP equipped with both cell-specific and UE-specific parameters. The UE may obtain the cell-specific and UE-specific parameters via different signaling messages. For example, during the initial access, the UE can obtain the cell-specific parameters of the DL/UL BWP #0 by reading the SIB1. The UE can further obtain the UE-specific parameters upon RRC configuration after the initial access. Option 2 is appealing in the deployments where multiple DL/UL BWPs are not needed. In this case, the network can set up a fully operational connection with a UE by only configuring DL/UL BWP #0 using Option 2.

NR supports configurations of up to four "RRC-configured" DL/UL BWPs. The DL/UL BWP #0 configured by Option 1 only has cell-specific parameters and is not counted as a "RRC-configured" BWP. Therefore, additional four DL/UL BWPs #1, #2, #3, and #4 may be consecutively configured. The DL/UL BWP #0 configured by Option 2 has both cell-specific and UE-specific parameters and thus is counted as a "RRC-configured" BWP. Therefore, additional three DL/UL BWPs #1, #2 and #3 may be consecutively configured. Figure 3 provides an illustration of the BWP configuration options in NR.

## IV. BANDWIDTH PARTS SWITCH

### A. RRC Reconfiguration Based Bandwidth Parts Switch

When more than one UE-specific DL/UL BWPs are configured to the UE on a serving cell, the first active DL/UL BWP, if configured, indicates the DL/UL BWP to be activated upon RRC (re-)configuration for an SpCell, and upon activation of an SCell. If the first active BWP is not configured, there is no BWP switch upon RRC (re-)configuration. The first active DL/UL BWP is always configured upon SCell addition, upon PCell change in MCG, and PSCell addition or change in SCG.

| # of RRC configured BWPs, excluding the initial BWP | Bitwidth of BWP indicator field | BWP indicator in DCI 0_1/1_1 | BWP Id | Comments | |
|---|---|---|---|---|---|
| 0 | 0 | Absent | 0 | For Option 1, DCI 0_1/1_1 is not applicable for operation on BWP#0. For Option 2, this is single BWP operation on BWP #0. | |
| 1 | 1 | 0, 1 | 0, 1 | Switch among BWP 0, 1 | For Option 1, once switch to BWP 0, cannot switch out by DCI |
| 2 | 2 | 00, 01, 10 | 0, 1, 2 | Switch among BWP 0, 1, 2 | |
| 3 | 2 | 00, 01, 10, 11 | 0, 1, 2, 3 | Switch among BWP 0, 1, 2, 3 | |
| 4 | 2 | 00, 01, 10, 11 | 1, 2, 3, 4 | Switch among BWP 1, 2, 3, 4 | |

**Table 1: Interpretation of BWP indicator field in DCI format 0_1/1_1 for DCI-based BWP switch.**

For BWP configuration Option 1, switch from the initial DL/UL BWP to another DL/UL BWP requires RRC reconfiguration since only DCI format 1_0/0_0 can be used with initial DL/UL BWP without dedicated configuration which does not support DCI-based BWP switch.

For RRC-based BWP switch, there is a delay of receiving (for DL active BWP switch) or transmitting (for UL active BWP switch) on the new BWP on the serving cell after the UE receives RRC reconfiguration involving active BWP switch or parameter change of its active BWP. The delay requirement for RRC-based BWP switch, within which UE shall complete the switch of active DL and/or UL BWP, is the sum of processing delay for RRC procedure and the delay for UE to perform BWP switch. The processing delay requirements for RRC procedure are in the range of 5-80 ms and differ among connection control procedures [9]. The delay requirement for UE to perform RRC-based BWP switch is a few milliseconds.

### B. DCI-Based Bandwidth Parts Switch

With initial DL/UL BWP and one or more additional DL/UL BWPs being configured to a UE, the network can schedule the UE to switch the active DL/UL BWP from one configured BWP to another using BWP indicator in DCI format 1_1/0_1. The possibility of DCI based BWP switch involving BWP #0 is dependent on BWP configuration option, as described in Table 1. DCI format 1_1 and DCI format 0_1 are non-fallback DCI formats for downlink assignment and uplink grant, respectively [10]. They support the full set of NR features and their fields are largely configurable. On the other hand, fallback DCI formats 1_0 and 0_0, used respectively for downlink assignment and uplink grant, do not contain BWP indicator field and thus do not support DCI-based BWP switch.

BWP field in DCI format 1_1/0_1 has the bitwidth of 0 – 2. The exact value is determined by the number of RRC configured DL/UL BWPs, excluding the initial DL/UL BWP. Table 1 provides the interpretation of BWP indicator field for DCI-based BWP switch.

There is a transmission/reception delay between network and UE associated with DCI-based BWP switch. UE shall complete the switch of active DL and/or UL BWP within the required BWP switch delay. BWP switch delay requirements are listed in Table 2 for both DCI- and timer-based BWP switch [11]. The switch delay denoted by $T_{BWPswitchDelay}$ for DCI-based BWP switch is defined as the slot offset between the DL slot in which the UE received switch request and the first slot in which the UE shall be able to receive PDSCH (for DL active BWP switch) or transmit PUSCH (for UL active BWP switch) on the new BWP. There are two levels of BWP switch delay requirement, type 1 and type 2, as given in Table 2. The UE is not required to transmit UL signals or receive DL signals during the time duration $T_{BWPswitchDelay}$ on the serving cell where DCI-based BWP switch occurs.

Note that the BWP switch delay is dependent on SCS. If the BWP switch happens between BWPs of different SCS values, the switch delay requirement is determined by the smaller SCS.

| SCS (kHz) | NR Slot length (ms) | BWP switch delay requirement $T_{BWPswitchDelay}$ (slots) | |
|---|---|---|---|
| | | Type 1 | Type 2 |
| 15 | 1 | 1 | 3 |
| 30 | 0.5 | 2 | 5 |
| 60 | 0.25 | 3 | 9 |
| 120 | 0.125 | 6 | 18 |

**Table 2: DCI- and timer-based BWP switch delay requirements.**

### C. Timer-Based Bandwidth Parts Switch

The network may configure a UE with a BWP inactivity timer and a default DL BWP on a serving cell. The default DL BWP is one of the DL BWPs configured to the UE and becomes the active DL BWP upon expiry of the inactivity timer. If no default DL BWP is configured, the default DL BWP is the initial DL BWP. As mentioned in Section II, for unpaired spectrum (TDD), a DL BWP and an UL BWP with the same indices are linked and switched together. Thus, a DL BWP is effectively a DL/UL BWP pair in this case.

The granularity of the timer is 1 ms (i.e., 1 subframe) for FR1 and 0.5 ms for FR2. When the timer is running, the UE decrements the timer at the end of each subframe for FR1 or at the end of each half-subframe for FR2. The values for the BWP inactivity timer have the range of 2 – 2560 ms. The maximum value for the BWP inactivity timer matches the maximum value of discontinuous reception (DRX) inactivity timer, which allows for a configuration that prevents the timer from expiring while the DRX inactivity timer is running.

A UE starts the BWP inactivity timer of a serving cell, if configured, when it activates a DL BWP other than the default DL BWP. A UE restarts the BWP inactivity timer of the serving cell when it decodes a DCI with downlink assignment for the active DL BWP in paired spectrum, or when it decodes a DCI with downlink assignment or uplink grant for its active DL/UL

BWP pair in unpaired spectrum. A UE shall start/restart the BWP inactivity timer when a PDCCH for DCI-based BWP switch is received. BWP inactivity timer can only be started or restarted when there is no ongoing random-access procedure associated with the serving cell.

For timer-based BWP switch, the BWP switch transition time duration is from the subframe/half-subframe for FR1/FR2 immediately after a BWP inactivity timer expires until the beginning of a slot where the UE can receive or transmit. The UE is not required to receive or transmit on the serving cell during the transition. Timer-based BWP switch shares the same BWP switch delay requirements as DCI-based BWP switch, as shown in Table 2.

Switching between configured BWPs may also happen when random-access procedure is initiated on a serving cell. UL BWP is switched to the initial UL BWP if the physical random access channel (PRACH) occasions are not configured for the active UL BWP of the serving cell. If the serving cell is SpCell, the active DL BWP needs to be switched to the one with the same BWP index as the active UL BWP.

## V. UE CAPABILITIES OF BANDWIDTH PARTS SUPPORT

UEs typically support only a subset of the specified radio access features due to implementation constraints and test limitation. The UE sends its capability parameters to the network and the network shall configure and schedule the UE accordingly. In this section, we describe the BWP related UE capabilities and the corresponding parameters [12][13].

As mentioned in Section II, a UE only receives PDCCH and PDSCH in an active DL BWP and transmits PUCCH and PUSCH in an active UL BWP per serving cell. It is mandatory for a UE to support the basic BWP operation of one RRC configured DL BWP and one RRC configure UL BWP.

For the initial access, a UE needs to perform cell search and downlink synchronization by detecting SSB and acquire SIB1 by decoding DCI transmitted in CORESET#0. The bandwidths of SSB and CORESET#0 may or may not be included in a DL BWP. For reducing UE implementation complexity, BWP bandwidth restriction can be applied to avoid RF tuning. With BWP bandwidth restriction, a DL BWP includes the bandwidths of SSB and CORESET#0 (if present) for PCell/PSCell and includes the bandwidth of SSB (if present) for SCell. It is mandatory for a UE to support the basic BWP operation with the BWP bandwidth restriction. UE support of BWP operation without the BWP bandwidth restriction is an optional capability.

Supporting bandwidth adaptation with more than one DL/UL RRC configured BWPs and switching among BWPs are optional. UE may support bandwidth adaptation with up to two or four RRC configured DL and/or UL BWPs with the same numerology per serving cell, or with up to four RRC configured DL and/or UL BWPs with different numerologies per serving cell.

RRC-based BWP switch is a default function supported by all UEs. DCI- and timer-based BWP switches, which enable efficient bandwidth adaptation, are applicable to the UE supporting more than one RRC configured BWPs. The UE can also report which of the two switch delay requirements listed in Table 2 it supports.

BWP is introduced to NR to support flexible bandwidth operation by decoupling the channel bandwidth of a carrier from the UE channel bandwidth. From the physical layer design perspective, the bandwidth of a BWP spans from 1 RB to 275 RBs, although BWP sizes smaller than the resource block group (RBG) size or the precoding resource block group (PRG) are not supported in Release 15 [14].

## VI. USE CASES OF BANDWIDTH PARTS

### A. Flexible Bandwidth Support

Basic bandwidth flexibility has been introduced since LTE by supporting multiple carrier bandwidths and enabling carrier aggregation, and is adopted in NR. A normal LTE device is required to transmit and receive on the full carrier bandwidth of the frequency band supported by the device. LTE machine type communication (LTE-M) and Narrowband Internet of Things (NB-IoT) have been developed to relax this constraint. There has been growing demand for higher bandwidth flexibility in NR due to several reasons:

- NR should support network operation in a much wider range of spectrum with wider carrier bandwidth than LTE.
- NR should support a wide range of services and applications. They may have different requirements on throughput, latency, and reliability.
- UE devices of different bandwidth capabilities should be supported in the same NR network.

Besides carrier aggregation, the configuration of BWPs is one of the main building blocks to meet the new requirements of bandwidth flexibility in NR, though the current BWP scheme is not as flexible as LTE-M/NB-IoT when it comes to narrow bandwidth and low UE complexity. With DL/UL BWPs, along with DL/UL UE-specific channel bandwidth, configured to a UE (see Figure 1), the reception and transmission bandwidths for the UE are decoupled from each other and decoupled from carrier bandwidth.

### B. UE Power Saving

Power efficiency is an important design consideration for UE. Several basic UE power saving schemes from LTE are adopted in NR [15], including wakeup-sleep management for adaptation to traffic load in time and fast activation/deactivation of SCell for adaptation to traffic load in frequency. Wakeup-sleep management such as connected mode DRX (cDRX) is beneficial for UE handling bursty data traffic by switching between network access mode and power efficient mode. Fast activation/deactivation of SCell helps UE to achieve power saving by adjusting bandwidth processing requirements at the granularity of component carrier level.

BWP-based bandwidth adaptation is introduced in NR to improve UE power efficiency by finer-granularity adaptation to traffic variation in frequency dimension [5]. Bandwidth adaptation is typically achieved by configuring the UE with multiple BWPs and dynamically switching the UE's active BWP among the configured BWPs. For maximizing UE power saving gain, BWP-based bandwidth adaptation is usually applied in conjunction with cDRX and/or fast activation/deactivation of SCell.

*C. Fast Change of UE Configuration*

From a UE's perspective, all physical channels and most physical signals are configured per BWP by the network. Switching among multiple BWPs for the UE usually occurs with UE configuration change as well. Each BWP has specific physical characteristics including frequency location, bandwidth, SCS, and cyclic prefix. UE configuration needs to, at least, convey the physical characteristics of the associated BWP. The network can also configure BWPs to a UE with the same (or similar) physical characteristics but with different UE configurations. For example, two BWPs with the same physical characteristics (e.g. same bandwidth, position, SCS) may be configured to a UE with different uplink waveforms: One BWP is configured with cyclic-prefix OFDM (CP-OFDM) waveform and the other BWP is configured with discrete Fourier transform spread OFDM (DFT-s-OFDM) waveform.

By applying DCI-based BWP switching among such BWPs the network may "reconfigure" the UE within $1 - 3$ ms (c.f. Table 2), which is faster, by at least one order of magnitude, than the legacy RRC reconfiguration procedure. DCI-based BWP switch for fast change of UE configuration is a complementary approach limited by the maximum of four RRC configured DL/UL BWPs in Release 15.

Other use cases and application scenarios can be derived based on BWP concept in NR. Network may provide services with different levels of quality of service (QoS) to the same UE or different UEs. BWPs configured with different configurations can be applied to accommodate different service requirements.

## VII. CONCLUSIONS

BWP is a basic concept in 5G NR. This article provides an overview of the essentials of BWP in the NR technical specifications, including the fundamental BWP concepts, BWP configuration methods, BWP switch mechanisms, and UE capabilities in terms of BWP support. As highlighted in this article, BWP may have the potential of enabling more flexible bandwidth support, reducing UE power consumption, achieving fast change of UE configuration, among others. As 5G rollout is happening, it will be interesting to see how BWP will be used in the real networks.